\documentstyle[12pt,epsfig]{article}
\begin{document}
\begin{center}
{\large \bf
Is there a unique thermal dilepton source in the reactions\\
Pb(158 A$\cdot$GeV) + Au, Pb?
} \\[6mm]
{\sc
K. Gallmeister$^a$, B. K\"ampfer$^a$, O.P. Pavlenko$^{a,b}$} \\[6mm]
$^a$Forschungszentrum Rossendorf, PF 510119, 01314 Dresden, Germany \\[1mm]
$^b$Institute for Theoretical Physics, 252143 Kiev - 143, Ukraine
\end{center}

\vspace*{9mm}

\centerline{Abstract} 
An analysis of
the dilepton measurements of the reactions Pb(158 A$\cdot$GeV) + Au, Pb
by the CERES and NA50 collaborations 
points to a unique
thermal source contributing to the invariant mass and transverse 
momentum spectra. \\[2cm]

{\bf Introduction:} 
Dileptons are penetrating probes which carry nearly undisturbed
information about early, hot and dense matter stages in relativistic
heavy-ion collisions. Some effort, however, is needed for
disentangling the various sources contributing to the observed
yields and for identifying the messengers from primordial states of
strongly interacting matter.
 
The dielectron spectra for the reaction Pb(158 A$\cdot$GeV) + Au measured by
the CERES collaboration \cite{CERES_1} cannot be described by a superposition
of $e^+ e^-$ decay channels of final hadrons, i.e.\ the hadronic cocktail.
A significant additional source of dielectrons must be there. Since the
data \cite{CERES_1} cover mainly the invariant mass range $M <$ 1.5 GeV the
downward extrapolation of the Drell-Yan process is not 
an appropriate explanation.
Also correlated semileptonic decays of open charm mesons have been excluded
\cite{PBM}. As a widely accepted explanation, a thermal source is found to
account for the data (cf. \cite{CERES_exp,Rapp_Wambach} and further
references therein, in particular with respect to in-medium effects and
chiral symmetry restoration).

Very similar, the NA50 collaboration has found, for the reaction
Pb(158 A$\cdot$GeV) + Pb, that the superposition of
Drell-Yan dimuons and open charm decays does not explain the data
in the invariant mass range 1.5 GeV $< M <$ 2.5 GeV \cite{NA50_1}.
Final state interactions \cite{Lin} or abnormal charm enhancement
\cite{NA50_1,NA50_2} have been proposed as possible explanations. Here we
try to explain the NA50 measurements by another idea \cite{Rapp},
namely a thermal source \cite{Gallmeister,Shuryak},
which was already found to
account for the data in the intermediate invariant mass range
in the reaction S (200 A$\cdot$GeV) + W 
\cite{th_source}.
We present a schematic model describing at the same time
the CERES and NA50 data.

{\bf The model:} 
Since we include the corresponding detector acceptances a good starting
point for Monte Carlo simulations is the differential dilepton spectrum
\begin{equation}
\frac{dN}{p_{\perp \,1} \, d p_{\perp \,1} \, 
p_{\perp \,2} \, d p_{\perp \,2} dy_1 \, dy_2 \,
d \phi_1 \, d \phi_2}
=
\int d^4 Q \, d^4 x \, \frac{dR}{d^4 Q \, d^4 x}
\delta^{(4)} (Q - p_1 - p_2),
\label{rate_1}
\end{equation}
where $Q = p_1 + p_2$ is the pair four-momentum,
$p_{1,2}$ are the individual lepton four-momenta composed of transverse momenta
$p_{\perp \,1,2}$ and rapidities $y_{1,2}$ and azimuthal angles
$\phi_{1,2}$. Here we extensively employ the quark - hadron duality
\cite{Rapp_Wambach,Rapp,th_source,duality} 
and base the rate $R$ on the lowest-order
quark - antiquark ($q \bar q$) annihilation rate 
(cf. \cite{KKMcLR,Ruuskanen})
\begin{equation}
\frac{dR}{d^4 Q \, d^4 x} = \frac{5 \alpha^2}{36 \pi^4} 
\exp \left\{ - \frac{u \cdot Q}{T} \right\},
\label{rate_2}
\end{equation}
where $u(x)$ is the four-velocity of the medium depending on the space-time
as also the temperature $T(x)$ does.
Note that, due to Lorentz invariance, $u$ necessarily enters this expression.
The above rate is in Boltzmann approximation, and a term related to the
chemical potential is suppressed. As shown in \cite{Rapp_Wambach} the
$q \bar q$ rate deviates from the hadronic one at $M <$ 300 MeV, but in this
range the Dalitz decays dominate anyhow; in addition, in this range the
thermal yield is strongly suppressed by the CERES acceptance.
In the kinematical regions we consider below, the lepton masses can be
neglected.

Performing the space-time and momentum integrations in 
eqs.~(\ref{rate_1}, \ref{rate_2})  one gets
\begin{equation}
\frac{dN}{d p_{\perp \,1} \, d p_{\perp \,2} 
dy_1 \, dy_2 \, d \phi_1 \, d \phi_2}
= 
\frac{5 \alpha^2}{72 \pi^5} p_{\perp \, 1} p_{\perp \, 2}
\int_{t_i}^{t_f} dt \, V(t) \, E,
\label{rate_3}
\end{equation}
\begin{eqnarray}
E &=& \left\{
\begin{array}{ll}   
\exp \left\{ -
\frac{M_\perp \cosh Y \cosh \rho (r,t)}{T(r,t)} \right\}
\frac{\sinh \xi}{\xi}
& \mbox{for}
\quad 3D,\\
K_0 \left( \frac{M_\perp \cosh \rho (r,t)}{T(r,t)} \right)
I_0 \left( \frac{Q_\perp \sinh \rho (r,t)}{T(r,t)} \right)
& \mbox{for}
\quad 2D,\\
\end{array}   
\right.  \\
V(t) &=& \left\{
\begin{array}{ll}  
4 \pi \int dr \, r^2
& \mbox{for}
\quad 3D,  \\
t \int dr \, r
& \mbox{for}
\quad 2D,  \\
\end{array}   
\right.
\end{eqnarray}
where $V(t)$ acts on $E$, and $3D$ means spherical symmetric expansion,
while $2D$ denotes the case of longitudinal boost-invariant and
cylinder-symmetrical transverse expansion; the quantity $\xi$ is defined as
$\xi = T^{-1} \sinh \rho \sqrt{M_\perp^2 \cosh^2 Y - M^2}$,
and $\rho(r,t)$ is the radial or transverse expansion rapidity;
$K_0$ and $I_0$ are Bessel functions \cite{KKMcLR}.
The components of the lepton pair four-momentum
$Q = (M_\perp \cosh Y, M_\perp \sinh Y, \vec Q_\perp)$
are related to the individual lepton momenta via
\begin{eqnarray}
M_\perp^2 
&=& p_{\perp \, 1}^2 + p_{\perp \, 2}^2 +
2 p_{\perp \, 1} p_{\perp \, 2} \cosh (y_1 -y_2), \\
\vec Q_\perp &=& \vec p_{\perp \, 1} + \vec p_{\perp \, 2},\\
M^2 &=& M_\perp^2 - Q_\perp^2,\\
\tanh  Y 
&=& 
\frac{p_{\perp 1} \, \sinh y_1 + p_{\perp 2} \, \sinh y_2 }
{p_{\perp 1} \, \cosh y_1 + p_{\perp 2} \, \cosh y_2 }. 
\end{eqnarray}
It turns out that the shape of the invariant mass spectrum
$d N / (dM \, dY \vert_{|Y| < 0.5} \, dt \, d V(t))$, which is determined only
by the emissivity function $E$, does not depend on the flow rapidity 
$\rho$ in the 2D case \cite{KKMcLR}, and in the 3D case 
for $T$ = 120 $\cdots$ 220 MeV and $\rho < 0.6$ there is also no
effect of the flow. The analysis of transverse momentum spectra of
various hadrons species point to an average transverse expansion velocity
$\bar v_\perp \approx$ 0.43 \cite{BK_2} at kinetic freeze-out,
while a combined analysis of hadron spectra
including HBT data yields $\bar v_\perp \approx$ 0.55 \cite{Heinz}.
Therefore, $\rho < 0.6$ is the relevant range for the considered reactions.
  
We note further that the invariant mass spectra 
$d N / (dM \, dY \vert_{|Y| < 0.5} \, dt \, d V(t))$
for the 3D and 2D cases differ only marginally.
Relying on these findings one can approximate the emissivity function
$E$ by that of a ''static'' source at midrapidity, as appropriate only for
symmetric systems,
\begin{equation}
E = \exp \left\{ - \frac{M_\perp \cosh Y}{T(t)} \right\},
\label{rate_4}
\end{equation}
thus getting rid of the peculiarities of the flow pattern.
We would like to emphasize the approximate character of
eq.~(\ref{rate_4})
because once cooling and dilution of the matter are included, they are
necessarily accompanied by expansion and flow. One has therefore to
check to which degree the flow affects the dilepton observables.

In contrast to the invariant mass spectra, 
the transverse momentum or transverse mass spectra are
sensitive to the flow pattern \cite{KKMcLR,Asakawa_Ko_Levai,BK_1}, in general.
A value of $\rho =$ 0.4, for example, causes already a sizeable change of the shape 
of the spectra $dN / (dQ_\perp dY \vert_{|y| < 0.5} \, dt \, d V(t))$ 
compared to $\rho = 0$, in particular in the large-$Q_\perp$ region. 
The differences between the 2D and 3D cases are not
larger than a factor of 2 and,
in a restricted $Q_\perp$ interval,
can be absorbed in a renormalization.
The most striking difference of the 2D and 3D scenarios is seen in the
rapidity spectrum: for 2D it is flat, while in the 3D case it is
localized at midrapidity (values of $\rho <$ 0.6 also do not change
the latter rapidity distribution). Below we shall discuss which space is left
to extract from the dilepton data in restricted acceptance regions hints for
the flow pattern when the other dilepton sources are also taken into account.

{\bf Comparison with data:} 
In line with the above arguments we base our rate calculations on
eqs.~(\ref{rate_3}, \ref{rate_4}) and use the parameterizations
\cite{Klingel_Weise}
\begin{eqnarray}
T &=& (T_i - T_\infty) \exp \left\{ - \frac{t}{t_2} \right\} + T_\infty,\\
V &=& N \exp \left\{ \frac{t}{t_1} \right\}.
\label{history}
\end{eqnarray}
with $T_i =$ 210 MeV, $T_\infty =$ 110 MeV, $t_1 =$ 5 fm/c, $t_2 =$ 8 fm/c,
$N = \frac{A + B}{2.5 n_0}$ with $A, B$ as mass numbers of the
colliding systems and $n_0 =$ 0.17 fm${}^{-3}$.
We stop the time evolution at $T_f =$ 130 MeV.

In fig.~1 we show the comparison with the preliminary CERES data applying the
appropriate acceptance \cite{CERES_1}. One observes a satisfactory overall
agreement of the sum of the hadronic cocktail and the thermal
contribution with the data.
It is the thermal contribution which fills the hole of
the cocktail around $M = 0.5$ GeV in the invariant mass distribution
in fig.~1a. In the mass bin $M =$ 0.25 $\cdots$ 0.68 GeV the thermal
yield is seen (fig.~1b) to dominate at small values of the transverse 
momentum $Q_\perp$. In this region of $Q_\perp$ transverse flow
effects are not important. The large-$Q_\perp$ spectrum is dominated
by the cocktail. For higher-mass bins the thermal
yield in the region of the first two data points is nearly 
as strong as the cocktail
and rapidly falls then at larger values of $Q_\perp$ below the cocktail. 
Therefore, the flow effects turn out to be of minor
importance for the present analysis, since within our framework the
transverse flow shows up at larger values of $Q_\perp$.

The question now is whether the same thermal source model 
accounts also for the NA50 data \cite{NA50_1}.
In the mass range $M > 1$ GeV, the Drell-Yan dileptons and dileptons 
from correlated semileptonic decays of open charm mesons must be included.
To get the corresponding yields for Pb + Pb collisions from 
PYTHIA \cite{PYTHIA} 
the overlap function $T_{AA} =$ 31 mb${}^{-1}$ is used.
We have carefully checked that our PYTHIA calculations with K
factors $K_{\rm DY} = 1.23$ and $K_{\rm charm} = 4$, adjusted to
Drell-Yan data \cite{DY_data} and identified open charm
data (cf. \cite{PBM} for a data compilation), 
and intrinsic transverse parton momentum
$\langle k_\perp \rangle = 0.9$ GeV
coincide with results
of the NA50 collaboration. In particular, we reproduce with
$\chi^2_{\rm d.o.f.} = 0.24$ the anticipated NA50 result \cite{NA50_1}
that the data of the most central collisions (except the $J/\psi$
region) are described by the Drell-Yan yield
$+$ 2.8 $\times$ the yield from correlated semileptonic decays of 
open charm. In this way we get some confidence in our acceptance
routine which essentially consists of geometrical cuts \cite{NA50_1}
and a suitable minimum single-muon energy of ${\cal O}$(11.5 GeV)
\cite{Shuryak}.

The resulting invariant mass spectra, including the thermal source 
contribution, are displayed in fig.~2a. The thermal source, with
strength adjusted by the above comparison with CERES data, is needed
to achieve the overall agreement with data. For $M < 2$ GeV the
thermal contribution dominates over the Dell-Yan and charm
contributions. The value of $\chi^2_{\rm d.o.f.} = 1.38$ quantifies
that the very details of the data are not perfectly described. This
may be attributed to the too schematic source model 
eq.~(\ref{rate_4}) and our approximate description of the more
involved NA50 acceptance. Nevertheless, the transverse momentum
spectrum for the mass bin 
$M =$ 1.5 $\cdots$ 2.5 GeV
is nicely reproduced by the sum of Drell-Yan, open charm and thermal
contributions with 
$\chi^2_{\rm d.o.f.} = 2$.
The thermal yield is strongest at not too large values of
$Q_\perp$ where transverse flow effects can be neglected.
Therefore, it seems that from present dilepton measurements the
transverse flow can hardly be inferred. However, a reduction of the
above uncomfortably large value of the phenomenological parameter
$\langle k_\perp \rangle$ could be partially compensated by transverse
flow.

In a previous version of this note \cite{Gallmeister}
we have confronted our model with the findings of the NA50
collaboration according to the recipe ''reconstructed data'' =
the Drell-Yan yield
$+$ 3 $\times$ the dilepton yield from decays of $D$ mesons and
found a much better agreement. This fact points to the need to employ
the full
acceptance matrix in attempting a decision on the best data interpretation. 

{\bf Summary and discussion:} 
In summary we have shown that a very simplified, schematic model for thermal
dilepton emission, with parameters adjusted to the CERES data, also
accounts for the measurements of the NA50 collaboration. 
Our study points to a common thermal
source seen in different phase space regions in the dielectron and
dimuon channels. This unifying interpretation of different measurements
has to be contrasted with other proposals  of explaining the dimuon
excess in the invariant mass region 1.5 $\cdots$ 2.5 GeV either by
final state hadronic interactions  or an abnormally large
open charm production. The latter one
should be checked experimentally by a direct measurement of $D$ mesons
as envisaged in the proposal \cite{NA50_2},
thus providing a firm understanding of various dilepton sources.

Due to the convolution of the local matter emissivity and the space-time
history of the whole matter and the general dependence on the flow pattern,
it is difficult to decide which type of matter (deconfined or hadron matter)
emits really the dileptons. 
Our model is not aimed at answering this question. Instead,
with respect to chiral symmetry restoration, we 
apply the quark - hadron duality as a convenient way to roughly
describe the dilepton emissivity of matter by a $q \bar q$ rate, 
being aware that higher-order QCD processes change this rate 
\cite{BK_1,Aurenche} (this might partially be
included in a changed normalization $N$). In further investigations 
a more microscopically founded rate together with 
a more detailed space-time evolution must
be attempted and chemical potentials controlling the baryon and pion densities
must be included. This has been accomplished in \cite{Shuryak} to a
large extent with similar conclusions as ours.

{\bf Acknowledgments:} 
Stimulating discussions with 
P. Braun-Munzinger,
W. Cassing, 
O. Drapier,
Z. Lin, 
U. Mosel,
E. Scomparin,
J. Rafelski,
J. Stachel, and
G. Zinovjev are gratefully
acknowledged.
We thank R. Rapp for extensive conversations on his NA50 acceptance
routine, enabling the comparison with data,
and for informing us on \cite{Shuryak} prior to publication.
O.P.P. thanks for the warm hospitality of the nuclear theory group
in the Research Center Rossendorf.
The work is supported by BMBF 06DR829/1 and WTZ UKR-008-98.

\newpage

\begin{figure}[t]
\centering
~\\[-1cm]
\psfig{file=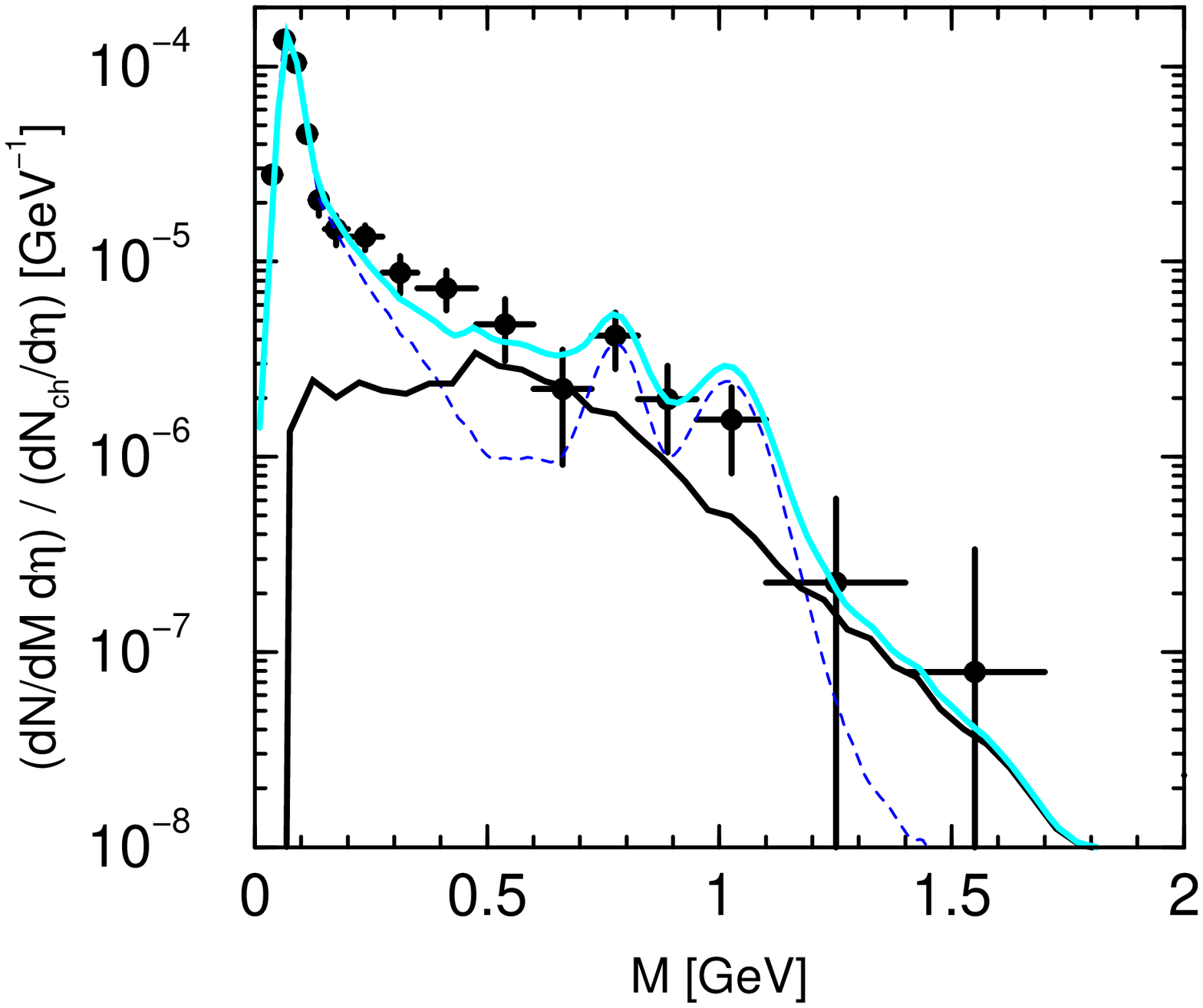,width=6.5cm,angle=-90}
\hfill
\psfig{file=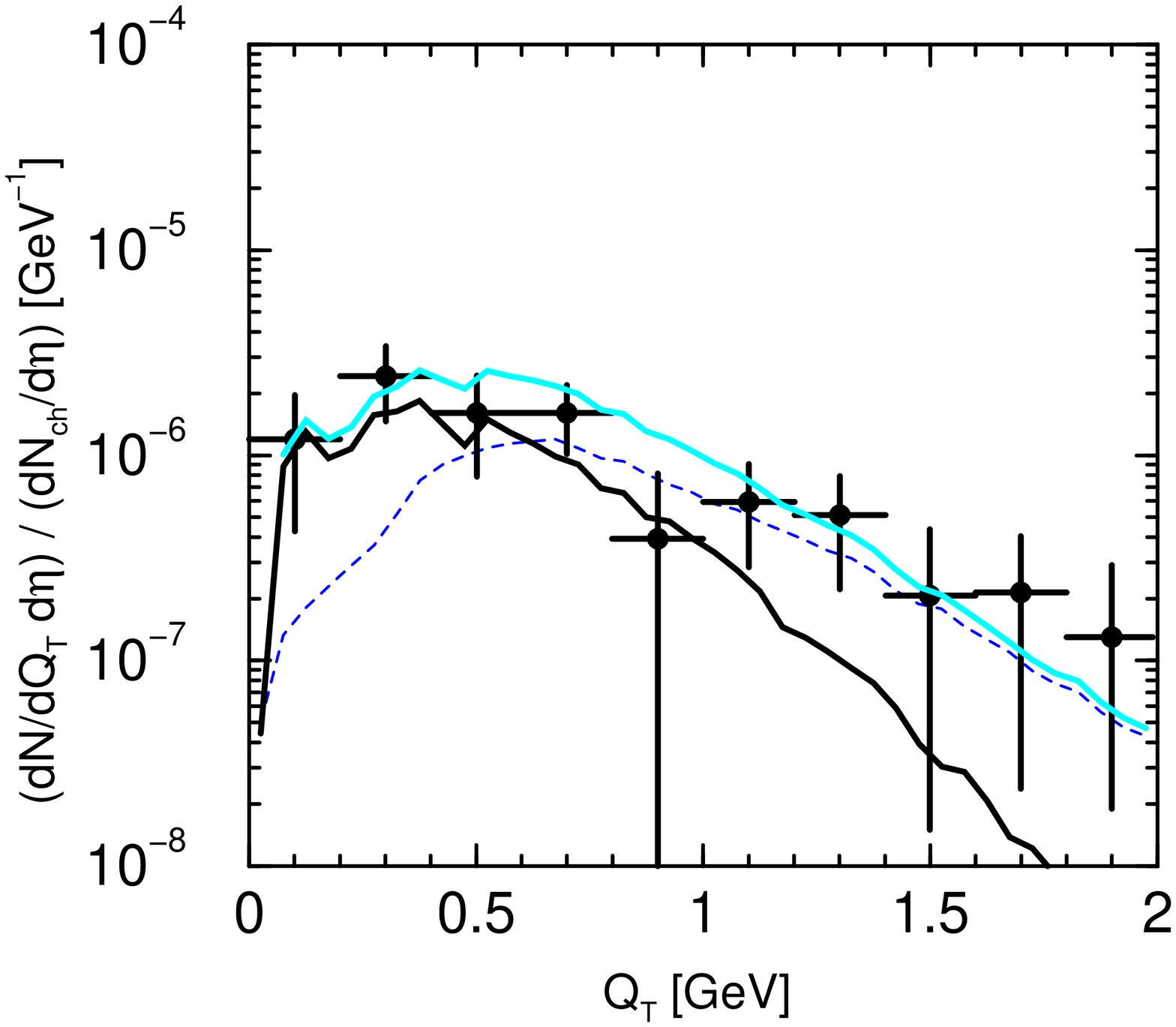,width=6.5cm,angle=-90}
~\\[.5cm]
\caption{
The preliminary CERES data \protect\cite{CERES_1}
and the hadronic cocktail \protect\cite{CERES_1} (dashed lines)
and the thermal yield (full curves). The sum
of the cocktail and the thermal yield is shown by the gray curves.
Left panel (a): the invariant mass spectrum
($\chi^2_{\rm d.o.f.} = 1.4$),
right panel (b): the transverse momentum spectrum for the mass bin
0.25 $\cdots$ 0.68 GeV
($\chi^2_{\rm d.o.f.} = .74$).
}
\label{fig.1}
\end{figure}

\begin{figure}[b]
\centering
~\\[-.1cm]
\psfig{file=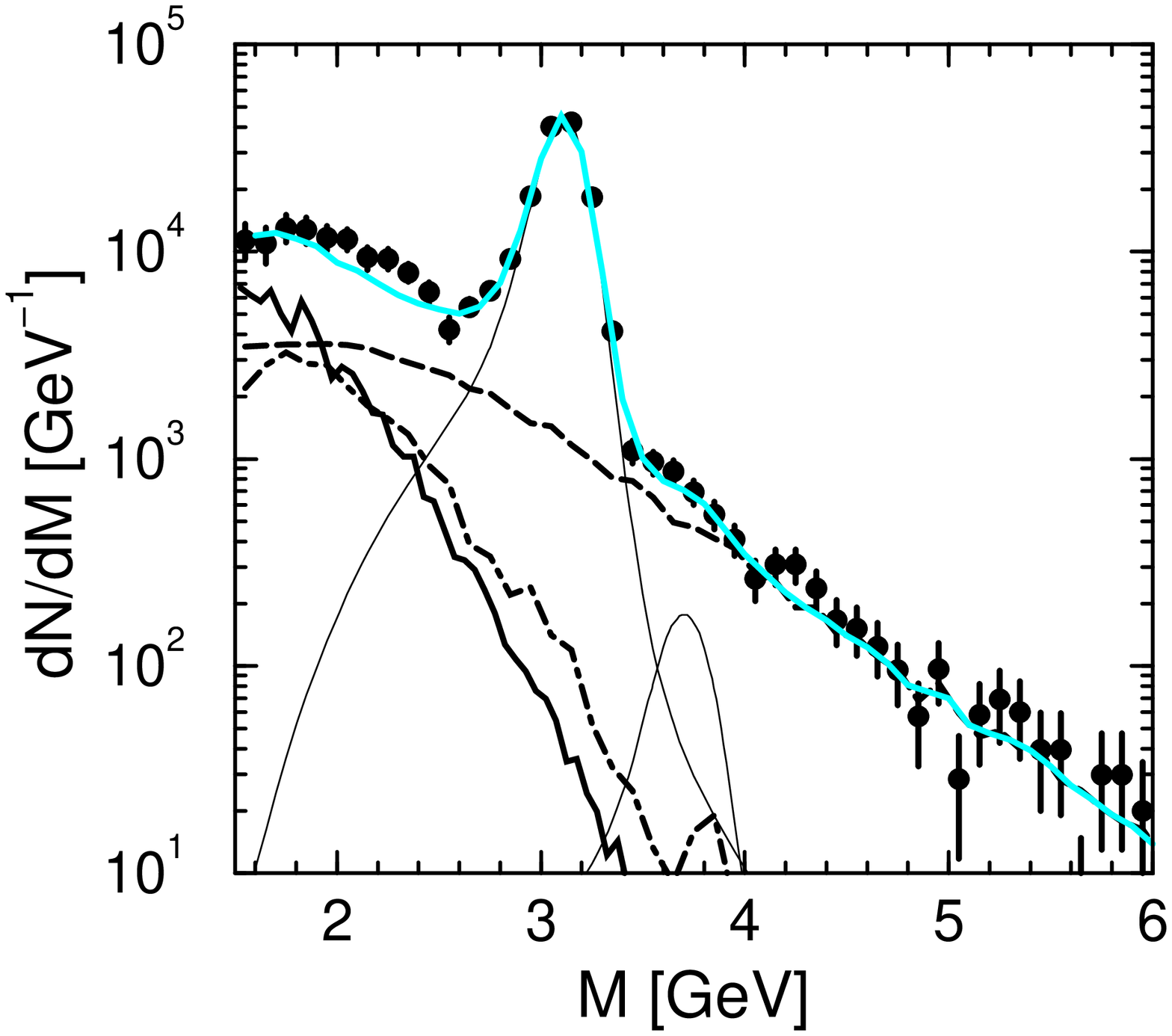,width=6.5cm,angle=-90}
\hfill
\psfig{file=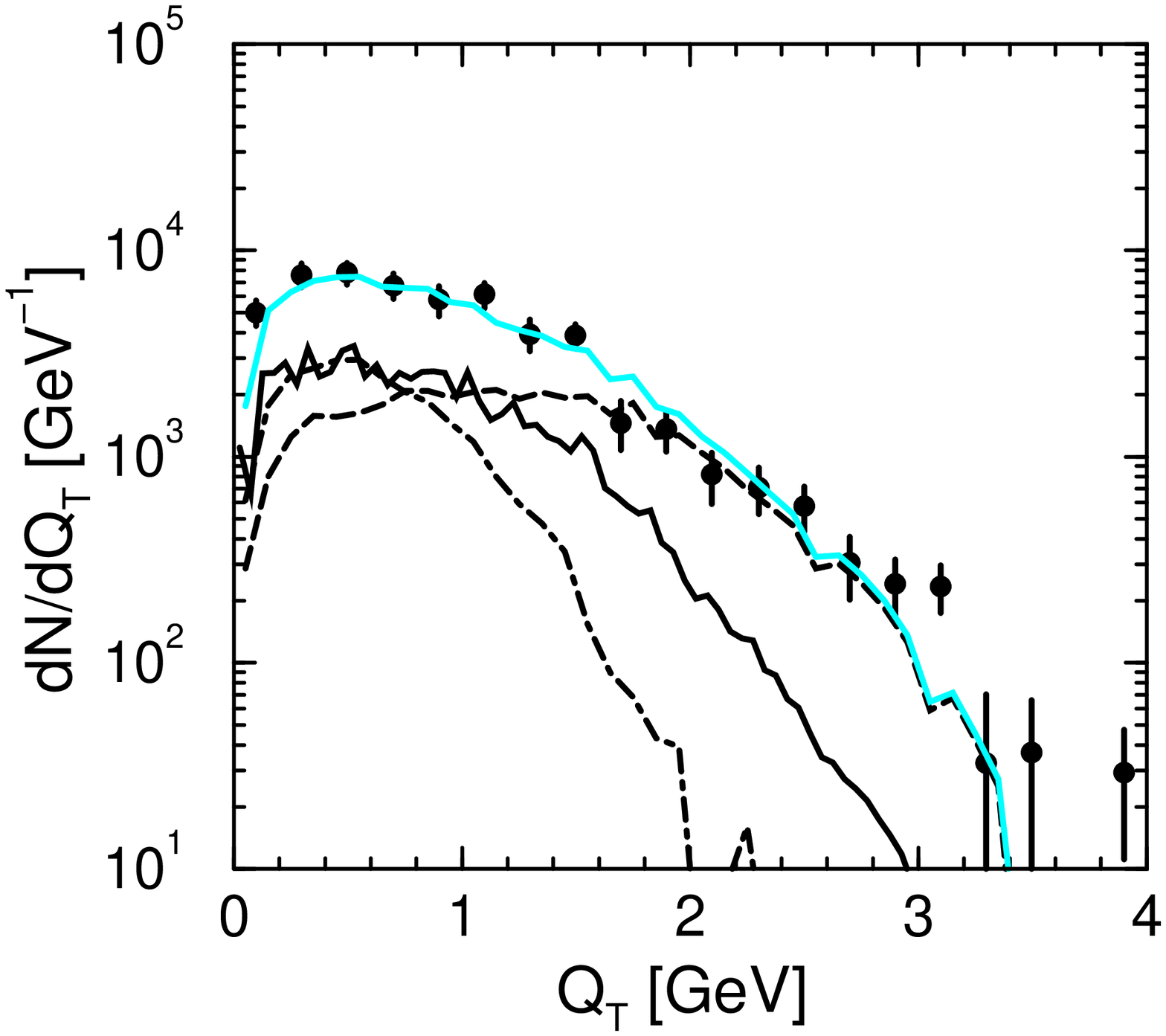,width=6.5cm,angle=-90}
~\\[.5cm]
\caption{
The preliminary NA50 data \protect\cite{NA50_1} 
in comparison with the thermal yield
(full curves), the Drell-Yan contribution (dashed curves) and the
contribution of open charm decays (dash-dotted curves). 
The $J/\psi$ and $\psi '$ curves (thin lines) are taken from 
\protect\cite{NA50_1}. The sum of
these contributions is displayed by the gray curves.
Left panel (a): the continuum invariant mass spectrum,
right panel (b): the transverse momentum spectrum for the mass bin
1.5 $\cdots$ 2.5 GeV.
}
\label{fig.2}
\end{figure}

\end{document}